\documentclass[sn-mathphys,Numbered,iicol]{sn-jnl}

\usepackage{graphicx}%
\usepackage{multirow}%
\usepackage{amsmath,amssymb,amsfonts}%
\usepackage{amsthm}%
\usepackage{mathrsfs}%
\usepackage[title]{appendix}%
\usepackage{xcolor}%
\usepackage{textcomp}%
\usepackage{manyfoot}%
\usepackage{booktabs}%
\usepackage{algorithm}%
\usepackage{algorithmicx}%
\usepackage{algpseudocode}%
\usepackage{listings}%

\newcommand{\tktable}[1]{\textsc{#1}{}}
\newcommand{\tkclass}[1]{\texttt{#1}{}}
\newcommand{\tkdir}[1]{\texttt{#1}{}}
\newcommand{\tkexec}[1]{\texttt{#1}{}}

\begin{document}

\title{The Toolkit for Nuclei library (TkN): a C++ interface to nuclear databases}

\author*[1]{\fnm{Jérémie} \sur{Dudouet}}\email{j.dudouet@ip2i.in2p3.fr}
\author[2]{\fnm{Diego} \sur{Gruyer}}

\affil*[1]{Universit\'e de Lyon 1, CNRS/IN2P3, UMR5822, IP2I, F-69622 Villeurbanne Cedex, France}
\affil[2]{Université de Caen Normandie, ENSICAEN, CNRS/IN2P3, LPC Caen UMR6534, F-14000 Caen, France}

\abstract{Over the past few decades, a vast amount of information on the structure of atomic nuclei has been collected, compiled, and evaluated. Accurate and reliable data are essential for the understanding of the behavior of atomic nuclei. Accessing and utilizing these data, spread among different databases, has remained challenging for many researchers due to the complexity and diversity of data formats. This article presents the Toolkit for Nuclei (TkN) C++ open-source library that provides easy access to nuclear structure data. This library is intended to be used in theoretical models, data analysis software or simulation codes. It utilizes a comprehensive database built from different official sources and frequently updated. The user interface allows to easily access and search for nuclear structure data and TkN can be compiled without any dependencies, facilitating its incorporation into various research projects.}

\maketitle

\section{Introduction}

As our understanding of nuclear physics continues to evolve, it is becoming increasingly important to have easy access to reliable, comprehensive and updated nuclear structure data. Accurate data are essential for theoretical model, data analysis software and simulation code development, and plays a crucial role in advancing our understanding of the behavior of atomic nuclei.

Over the past few decades, significant progress has been made in the experimental measurement of nuclear structure data, resulting in a vast amount of information on isotope ground states (binding energy, abundance...), excited states (energy, lifetime, spin-parity...), and decays (energy, branching ratio, multipolarity...). These information are compiled by different research groups and spread among different databases. Nuclear structure and decay data are, for example, compiled and evaluated by the International Network of Nuclear Structure and Decay Data (NSDD) evaluators that provide recommended values to be used in basic and applied research~\cite{nsdd}. 

These databases, already widely used in the scientific community, were historically created as text files with non-standard formats, and are mainly used to feed websites. However, accessing and utilizing these data with automatic processes, inside analysis codes, has remained challenging for many researchers due to the complexity and diversity of formats of the available data sets. To address this issue, we developed the Toolkit for Nuclei (TkN) C++ library that provides easy access to nuclear structure data. Our software utilizes a comprehensive and frequently updated database, built from different sources described in section~\ref{data-sources}.



\section{The TkN database}

The TkN library aims at giving access to a large amount of properties of chemical elements, nuclear isotopes, excited states and decays. These properties are gathered from different official sources and merged in a home-made and periodically updated SQlite~\cite{sqlite} database. This update process is guaranteed at least as long as the data sources are publicly available.

\subsection{Data sources}\label{data-sources}

The information on chemical elements are downloaded from the PubChem website~\cite{pubchem} in JSON format~\cite{json}. The different units were added manually and column names have been slightly adapted. X-ray data are extracted from the X-RAY DATA BOOKLET~\cite{xray-booklet} and then converted in JSON format. 

The largest database in nuclear physics that have been used in this project is provided by the International Network of Nuclear Structure and Decay Data evaluators (NSDD)~\cite{nsdd}, named "Evaluated Nuclear Structure Data File" (ENSDF)~\cite{ensdf}. It contains all the published values in nuclear physics on the excited states and decay properties, after being evaluated by the NSDD members using well-defined methodologies. The coordination of the evaluation effort and its web and journal dissemination is carried out by the National Nuclear Data Center (NNDC) at Brookhaven National Laboratory~\cite{NNDC}.

This database has the great advantage of having, for each nucleus, separated data-sets dedicated to the reaction mechanism that has been used to produce the nucleus of interest, and a combined one, named "adopted levels and gammas", that merge all the individual data-sets in one. The disadvantage of this database is that it is stored in an old ASCII-file that is not trivial to decode. NNDC is also providing another database in the same format, named "eXperimental Unevaluated Nuclear Data List" (XUNDL)~\cite{xundl}, containing published data, but still waiting to go through the NSDD evaluation process.

Another rich and useful database that has been used is the one distributed by the Nuclear Data Section of the International Atomic Energy Agency (IAEA-NDS), called Live Chart~\cite{iaea-livechart}. This database is using the ENSDF as input, but also a wider range of data sources very useful for nuclear ground state properties as:
\begin{itemize}
    \item the atomic mass evaluation~\cite{AME_2021,AME_2_2021} for mass excess values,
    \item the nubase evaluation~\cite{NUBASE_2021} for abundance and year of discovery of isotopes,
    \item the nuclear ground state charge radii~\cite{ANGELI201369},
    \item the nuclear magnetic dipole moments~\cite{Stone_2020_Magnetic,Stone_2023_Magnetic},
    \item the nuclear electric quadrupole moments~\cite{Stone_2023_Electric}
\end{itemize}

It provides powerful searching and plotting tools for web applications and recently, an API has been distributed to download the database content in CSV file format~\cite{iaea-api}. The disadvantage of this database is that it does not contains the individual data-sets of ENSDF (only the "adopted" one).


The last database that has been used in this project is the NUDAT3~\cite{nudat} database, distributed by NNDC. Most of its data sources are similar to the already discussed one, but some additional information can be obtained like the quadrupole deformation, or yields of $^{252}$Cf spontaneous fission source or $^{235}$U and $^{239}$Pu thermal neutron-induced fission. This database is directly downloaded from the NUDAT3 web site in JSON format.

As a consequence, it has been decided to combine these three databases to take all the data related to nuclear and decay properties from ENSDF and XUNDL databases, the ground state properties from the IAEA-NDS API and the remaining missing from NUDAT3. 

The TkN database is thus built by parsing ASCII, CSV and JSON files, with dedicated classes (see section \ref{dbbuilder}), to fill the home-made SQlite TkN database, much more efficient in terms of access-time. This SQlite database is to be seen as a TkN internal object, transparent for the user, used to fasten and facilitate the access to the data by the user interface (see \ref{user-interface}). It should be noted that TkN is not to be compared to the existing databases mentioned. It is a complementary utility for different applications. For a fast web search on a very specific query, the existing web-based databases like ENSDF or Live Chart are much more user-friendly and efficient. But for integration into analysis or simulation code with high computation needs, the TkN database is more suited.

\subsection{References}\label{data-references}

When a work using TkN is published, it is essential that, in addition to this article, references are made to the original databases used. If the experimental data used are from ENSDF or XUNDL databases, citation guidelines are specified on the ENSDF website~\cite{ensdf}. For data coming from NUDAT3, as mentioned on their website, it is recommended to use the following reference: ``\textit{National Nuclear Data Center, information extracted from the NuDat database, https://www.nndc.bnl.gov/nudat/}". For the other sources (chemical element properties, X-rays, or data coming from the IAEA-NDS API, the aforementioned references in section~\ref{data-sources} have to be used.

\subsection{Database structure}\label{data-structure}

\begin{figure*}[ht]
\begin{center}
\includegraphics[width=\linewidth]{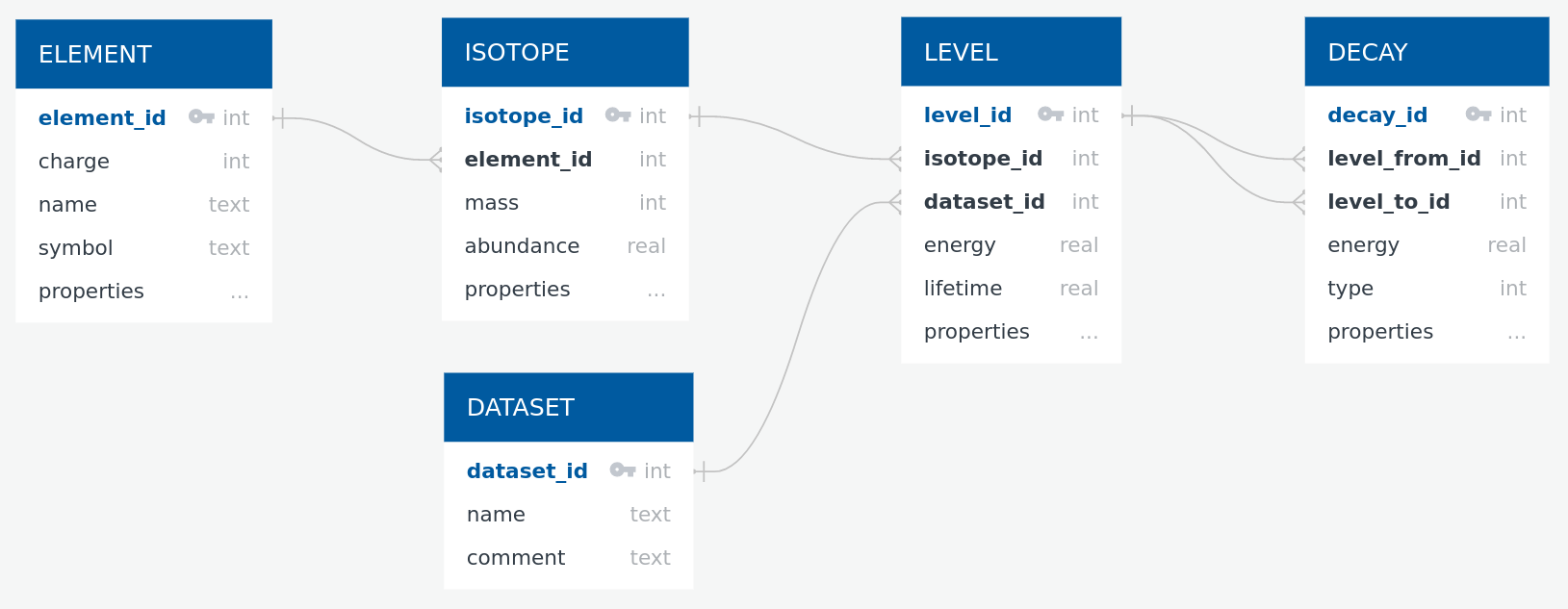}
\caption{\label{fig-layout} Schematic layout of the TkN database. The detailed content of each table is listed in \ref{app-tables}.}
\end{center}
\end{figure*}

The database structure was designed to avoid as much information redundancy as possible. It is composed of five tables linked with foreign joints as presented on figure~\ref{fig-layout}. The content of the tables is the following:

\begin{itemize}
    \item \tktable{element}: properties of chemical elements,
    \item \tktable{isotope}: ground state properties of all known isotopes of each element,
    \item \tktable{dataset}: a dataset links the information of a given nuclear level scheme (levels and decays) to the way the nucleus has been produced,
    \item \tktable{level}: excited states properties, each level being associated to the isotope it belongs and the dataset it was extracted from,   
    \item \tktable{decay}: decay  properties, associated to the parent and daughter levels.
\end{itemize}

The full list of table columns is detailed in appendix \ref{app-tables}. In its current format, the TkN database contains 118 elements, 3559 isotopes, 23523 datasets, 602385 levels and 760888 decays for a total size of about 180~MB.

For the data following the ENSDF data format~\cite{ensdf-format} (ENSDF and XUNDL sources), the choice has been made to only parse properties having a dedicated field in this format. Some other properties are given as comments in this format, as band structures, cross sections... Extracting automatically such information is not straight-forward and is thus not included in the current TkN version. Nevertheless, if users have specific needs that can be extracted, the missing data can be added case by case in the next TkN releases.

\subsection{Database generation}

X-ray and element properties are not frequently updated: the associated input files are directly stored in the TkN sources. However, isotope properties (ground state, levels and decays) are in constant evolution. For this reason, the TkN database is re-generated every month to take into account each new release of nuclear structure data evaluation.

To allow for data versioning, each database contains the dates of the ENSDF and XUNDL release used. The automatic database generation is handled by a \textit{crontab} task~\cite{crontab} executed on CNRS servers every 15th of each month. The user can then download or update the TkN database to integrate any new published physics properties, using the dedicated executable (see \ref{programs}).

\section{Code structure}

The TkN library is made of two parts : the first one allows to build the SQlite database, while the second one provides an interface for the user to access the data. Each part is made of C++ classes, under the \tkclass{tkn} namespace, and some program utilities.

\subsection{Database builder}\label{dbbuilder}

The first part of the TkN library is in charge of collecting the information from the different sources and building the SQLite~\cite{sqlite} database. From the code point of view, this corresponds to the classes in the directory \tkdir{src/builder} and of some executable programs dedicated to this task.

The \tkclass{builder} directory is composed of individual classes, each dedicated to its specific SQL table in the database. These classes are in charge of parsing the input data files from the different sources described in section \ref{data-sources}, and writing the extracted information in the database. The \tktable{dataset}, \tktable{level} and \tktable{decay} tables are handled by a global builder class (\tkclass{ensdf\_builder}) since they require to parse ENSDF data file format~\cite{ensdf-format}. 

Since the TkN database is not built by the user, this part of the library was only included in the project for sake of information. It is not compiled by default but can be activated using a dedicated CMake~\cite{cmake} flag.

\subsection{User interface}\label{user-interface}

The TkN user interface is a C++ shared library providing multiple classes to access the database and manipulate in an intuitive way its content. When creating a new nucleus (\tkclass{tknucleus} class), either by its name or atomic and mass number, the user can access any of the properties cited in section \ref{data-structure}, using specific predefined methods.

Data on excited states and decays are handled by the \tkclass{tklevel\_scheme} class. The ENSDF database provides information from different datasets that correspond to different reaction mechanisms used to produce a nucleus. To take it into account in the TkN database, each nucleus is associated to a list of available datasets, corresponding to a specific level scheme that contains its nuclear levels and decays. As defined in the ENSDF format, a global dataset named: "ADOPTED LEVELS, GAMMAS" is used to merge all the individual datasets in one. If it exists, this is the default dataset of any nucleus. To access to the nuclear properties of a nucleus, the user can  either use the global dataset, or select a specific dataset, depending on the physics case. In case of XUNDL datasets, a specific tag is added to specify that the dataset has not yet been evaluated.

As stated on the NNDC website~\cite{NNDC}, this database contains evaluated nuclear data - i.e. recommended values following a careful analysis of the available data. However, the accuracy of the data, the absence of errors or the absence of conflicting datasets cannot be guaranteed. The aim of this work is not to analyse the accuracy of the sources used, but to facilitate access to them.

The level scheme class (\tkclass{tklevel\_scheme}) also links levels and decays: each decay is associated to the parent and daughter levels, and each level contains a list of populating and depopulating decays, allowing for example to easily determine coincident decays.

The information on a measure (e.g. a level lifetime) is handled by a \tkclass{tkmeasure} object. A TkN measure contains, if defined : a value, an uncertainty (symmetric or asymmetric), a unit, and a tag. The tag can contain different information, such as the fact that a value is given as a lower or upper limit or results from a calculation or a systematic. For example, in the case of the spin and/or parity assignment of a nuclear level, its attribution can be tagged as firm or tentative. 

A measure comes, in most cases, with a unit. A dedicated unit manager has been implemented in TkN to allow a measure to be expressed in the desired unit. In general, units can be converted only in the scale of a same type (time, energy or length). One exception is given for the energy to time conversion. Indeed, the lifetime of very short lived levels is often expressed in terms of its intrinsic width (energy type unit).

The interface between the SQlite database and the user classes is handled by the \tkclass{tkmanager} class. For optimizing the TkN efficiency, it has been decided to pre-load all the element and nucleus ground state properties at the creation of the \tkclass{tkmanager} singleton. This allows the manager to know the full list of available nuclei. Methods are provided to allow for the user to conveniently loop over the available nuclei.
However, the amount of data contained in ENSDF, taking into account all the known levels, decays, and datasets is so important that is has been decided to not pre-load all of them by default. It has been preferred to load in memory the content of a level scheme (so for a given dataset), only when the user accesses it for the first time. Once a level scheme has been loaded, it stay in memory.

\subsection{TkN programs}\label{programs}

To generate the database, the TkN builder part comes with three executable programs: 
\begin{itemize}
    \item \tkexec{tkn-ensdf-update}: downloads the NNDC data files,
    \item \tkexec{tkn-create}: builds the TkN SQlite database, calling the different \tkdir{builder} classes,
    \item \tkexec{tkn-explorer}: explores the TkN SQlite database (a more user-friendly equivalent is included in the user interface programs).
\end{itemize}

Regarding the user interface, TkN provides different executable programs:
\begin{itemize}
    \item \tkexec{tkn-db-update}: downloads/updates TkN database,
    \item \tkexec{tkn-root}: opens a ROOT~\cite{root} prompt terminal with the TkN library linked (compiled only when using ROOT, see part \ref{prerequisites}),
    \item \tkexec{tkn-print}: lists in the terminal different properties for a given nucleus,
    \item \tkexec{tkn-config}: provides information on the current TkN installation (version, git branch, compilation flags...) that are useful for linking TkN to other packages (inspired from the \tkexec{root-config} executable of the ROOT package).
\end{itemize}

\subsection{Examples}

\begin{figure}[ht]
\includegraphics[width=\linewidth]{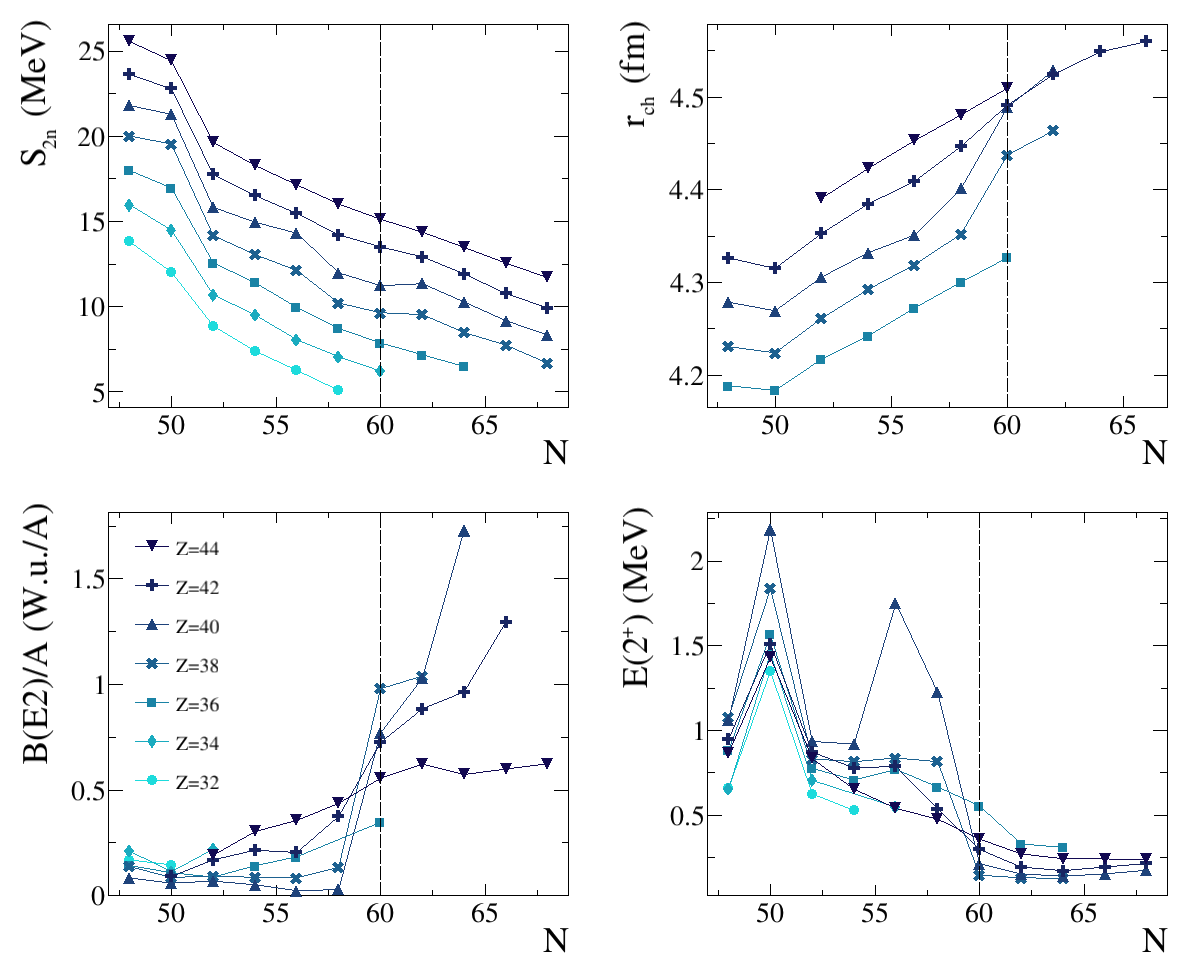}
\caption{\label{fig-def} Representation the sudden increase of nuclear deformation at N=60 in the A$\sim$100 island of deformation region, using four different experimental observables: the two neutron separation energy $S_{2n}$, the nuclear charge radius $r_{ch}$, the normalized electric transition probability B(E2:$2^+_1 \rightarrow 0^+_1$)/A, and the first $2^+$ state energy $E(2^+)$}
\end{figure}

A list of examples is provided in the TkN project to help users manipulate the various data accessible in the TkN database. As an example, the figure \ref{fig-def} has been produced with TkN and the source code is provided in the examples. This image represents the sudden increase of nuclear deformation at N=60 in the A$\sim$100 island of deformation region, using four different experimental observables: the two neutron separation energy, measured from mass measurements $S_{2n}$, the nuclear charge radius $r_{ch}$, the normalized electric transition probability of the first E2 gamma transition B(E2:$2^+_1 \rightarrow 0^+_1$)/A, and the first $2^+$ state energy $E(2^+)$.
Without TkN, producing such an image would require to extract data from the literature on ground state, nuclear levels and gamma decay properties for all these nuclei. Using TkN, it requires only a few lines of code.

\begin{figure}[ht]
\begin{center}
\includegraphics[width=\linewidth]{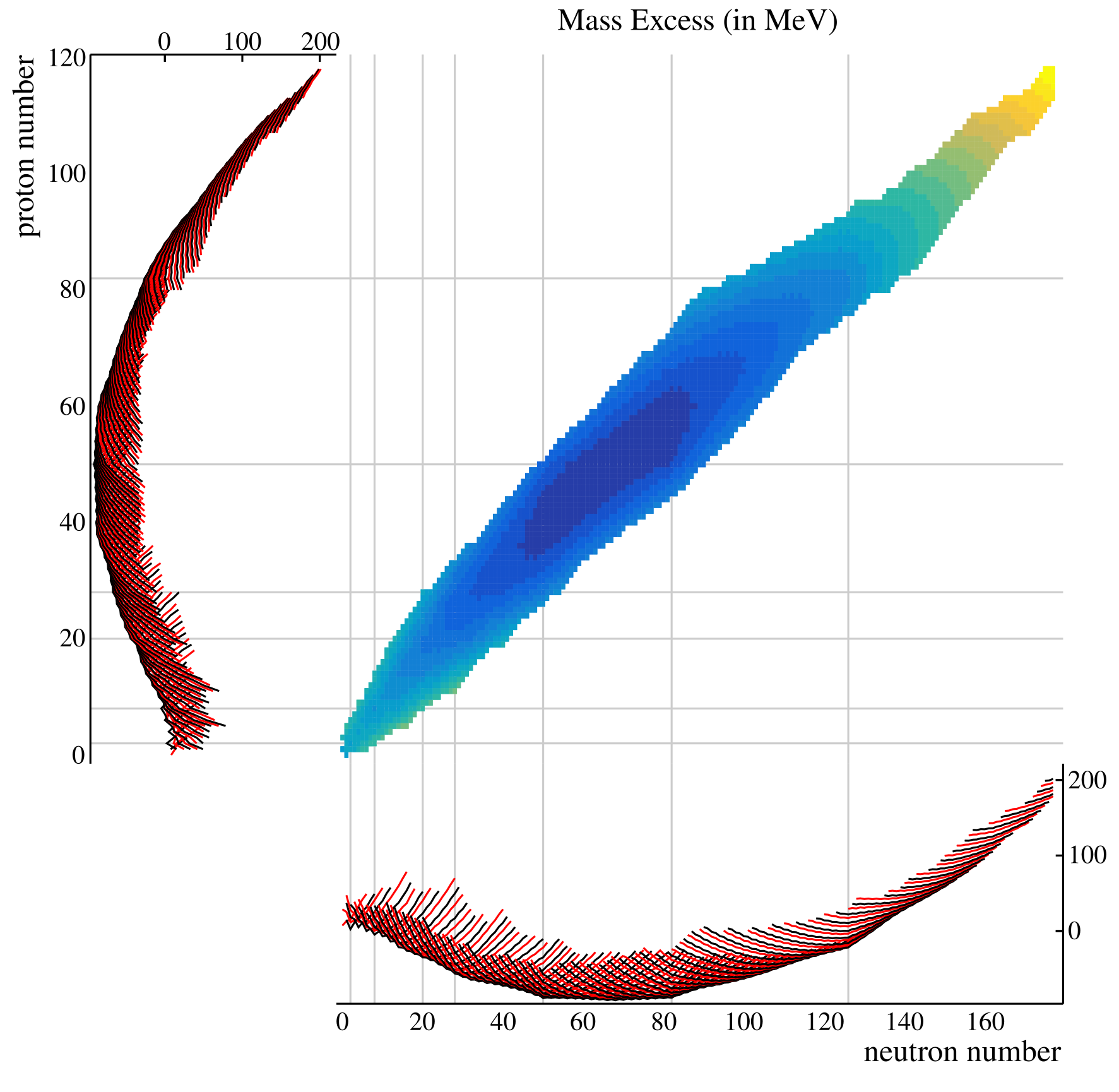}
\caption{\label{fig-me} Mass excess of all measured isotopes using the \tkclass{tknuclear\_chart} class for graphical representation.}
\end{center}
\end{figure}

In addition, a dedicated ROOT based C++ class, named \tkclass{tknuclear\_chart}, has been implemented to provide for the user a convenient way to represent physical values over the Segrè chart. Figure \ref{fig-me} is using this class to represent the nuclear mass excess as a function of the proton and neutron number. The bottom and left panels present the mass excess projected respectively on the neutron/proton numbers, for each isotopic/isotonic chains.

\subsection{Performances}

Building the \textit{tkmanager} (i.e. reading the full element and isotope tables) requires few milliseconds. The time needed to read all the levels and decays of all the known isotopes for the first time is about 13 seconds. As it is then stored in memory, a second reading of the complete database is of the order of 20 milliseconds.

The TkN user interface has been developed in order to be thread safe. In other words, it can be used in a multi-threaded program without any thread conflicts. Some tests have been realized on a machine with two CPU Intel(R) Xeon(R) Silver 4112 @ 2.60GHz, for a total of 16 available threads. The program used (available on the TkN documentation, see section~\ref{doc}) generates a random existing nucleus, from which a random nuclear level is read in its default level scheme. The time needed to generate $10^{8}$ levels with a single thread is around 170 seconds (13s to load the database and 157s for the processing). With 16 threads, this time drops to 30 seconds (still 13s to load the database and 17s for the processing).

\section{Software management}

\subsection{Software prerequisites}\label{prerequisites}

One of the main goal of the TkN library development was to provide a light library, easily linkable to any other software. For this reason, it has been decided to limit as much as possible the prerequisites:
\begin{itemize}
    \item an Unix architecture (MacOS or Linux),
    \item git (to download and update the project),
    \item a cmake version at least 3.8 (project written using the modern CMake techniques and methods),
    \item a compiler supporting C++17 (a C++11 patch is also provided),
    \item if linked with ROOT, a ROOT version greater than 6.20.
\end{itemize}

The SQLite library is integrated inside the TkN source code using the amalgamation file. The JSON parser NLohmann~\cite{lohmann} is also integrated as a single header file integration.

\subsection{Code distribution}

The TkN source code is hosted on the IN2P3 Gitlab server~\cite{tkn-git}. This allows for code version control, multiple developers collaboration, use of continuous integration process and automatic documentation generation.


Regarding the code development organisation, TkN is hosted on a main git repository containing a production and a pre-production branch. Each new development, that has been tested and validated is pushed on the pre-production branch. For each new code release, the pre-production branch is pushed on the production branch, generating a dedicated tag associated to a new TkN version number. Any developer works on its own forked repository, and merge the new developments on the main repository once validated.

The TkN version number is composed of a major and a minor number (e.g. 1.0). Each new minor value corresponds to new features in the library that don't break the compatibility with TkN database content. On the contrary, a new major number means that the database content has been modified (e.g. new physics properties added). For this reason, the database generation is associated to a TkN version number, to ensure the compatibility between the source code and the database format. 

This project is distributed under the CECILL-B license~\cite{cecill-b}, which is a free software license agreement for CNRS members.

\subsection{Continuous integration}

The source code is automatically built and tested using the Gitlab continuous integration (CI) capabilities. A dedicated docker~\cite{docker} image has been created to manage the CI operations. After any new git push on the TkN repository, the project is compiled with the different compilation options (e.g. with or without ROOT), and the unitary tests performed using the GoogleTest~\cite{googletest} framework are executed.

On the pre-production branch, a Cppcheck analysis~\cite{cppcheck} of the code is executed, and analyzed using SonarQube~\cite{sonarqube}.

\subsection{Documentation}\label{doc}

The TkN code is fully commented and documented on a dedicated web site~\cite{tkn-doc} hosted on Gitlab pages. The C++ code comments are integrated in the documentation using the Doxygen framework~\cite{doxygen}, providing indications on how to use any C++ methods and classes of the library.
In addition, the documentation web site includes a comprehensive user guide to provide users with information on data sources, latest release notes, help with installing TkN or linking it to other projects, and a comprehensive manual of TkN's main features. Finally, many examples are provided.

The TkN documentation is automatically generated by the Gitlab CI configuration. The home page of the web site is dedicated to the production branch of the Gitlab repository, but the other git branches documentation can be accessed using specific sub pages.

\section{Conclusions and perspectives}

The TkN project provides to physicists a new, fast and user friendly interface library to access and manipulate a huge amount of published data related to atomic and nuclear physics. As an example, it is already used in the Cubix software~\cite{cubix-userguide} that provides a new graphical user interface for gamma-ray spectroscopy analysis. TkN is used in this project to show, on a gamma-ray spectrum, the transitions already known in the observed nuclei, or to find which nucleus can correspond to the transitions measured in coincidence.

Nevertheless, a selection has been made by the developers, that is not exhaustive, of the data type to include in this database. Its content will evolve, depending on the uses and wishes of the community. For example, it is already planned for the next major version, to include particle non-electromagnetic decays. There are also longer-term plans to integrate nuclear reaction databases like the Experimental Nuclear Reaction Data (EXFOR)~\cite{exfor,exfor-web}.

Another aspect of the perspectives for the TkN library is to make it accessible to a wider part of the scientific community, by providing python and julia interfaces.

\section*{Acknowledgements} 

The authors thank Donnie Mason and Elizabeth Mccutchan for their help in decoding the ENSDF format, and Marco Verpelli for his help in using the LiveChart public API. More generally, the authors would like to thank the members of the NSDD for their intensive work on the evaluation of nuclear structure and decay data, and the IAEA for providing the LiveChart public API. The authors also thank Olivier Stezowski for useful discussions and providing inspirations with the GammaWare software. 


\section*{Data Availability Statement} 

No Data associated in the manuscript

\bibliography{TkN.bib}


\begin{thebibliography}{34}
\ifx \bisbn   \undefined \def \bisbn  #1{ISBN #1}\fi
\ifx \binits  \undefined \def \binits#1{#1}\fi
\ifx \bauthor  \undefined \def \bauthor#1{#1}\fi
\ifx \batitle  \undefined \def \batitle#1{#1}\fi
\ifx \bjtitle  \undefined \def \bjtitle#1{#1}\fi
\ifx \bvolume  \undefined \def \bvolume#1{\textbf{#1}}\fi
\ifx \byear  \undefined \def \byear#1{#1}\fi
\ifx \bissue  \undefined \def \bissue#1{#1}\fi
\ifx \bfpage  \undefined \def \bfpage#1{#1}\fi
\ifx \blpage  \undefined \def \blpage #1{#1}\fi
\ifx \burl  \undefined \def \burl#1{\textsf{#1}}\fi
\ifx \doiurl  \undefined \def \doiurl#1{\url{https://doi.org/#1}}\fi
\ifx \betal  \undefined \def \betal{\textit{et al.}}\fi
\ifx \binstitute  \undefined \def \binstitute#1{#1}\fi
\ifx \binstitutionaled  \undefined \def \binstitutionaled#1{#1}\fi
\ifx \bctitle  \undefined \def \bctitle#1{#1}\fi
\ifx \beditor  \undefined \def \beditor#1{#1}\fi
\ifx \bpublisher  \undefined \def \bpublisher#1{#1}\fi
\ifx \bbtitle  \undefined \def \bbtitle#1{#1}\fi
\ifx \bedition  \undefined \def \bedition#1{#1}\fi
\ifx \bseriesno  \undefined \def \bseriesno#1{#1}\fi
\ifx \blocation  \undefined \def \blocation#1{#1}\fi
\ifx \bsertitle  \undefined \def \bsertitle#1{#1}\fi
\ifx \bsnm \undefined \def \bsnm#1{#1}\fi
\ifx \bsuffix \undefined \def \bsuffix#1{#1}\fi
\ifx \bparticle \undefined \def \bparticle#1{#1}\fi
\ifx \barticle \undefined \def \barticle#1{#1}\fi
\bibcommenthead
\ifx \bconfdate \undefined \def \bconfdate #1{#1}\fi
\ifx \botherref \undefined \def \botherref #1{#1}\fi
\ifx \url \undefined \def \url#1{\textsf{#1}}\fi
\ifx \bchapter \undefined \def \bchapter#1{#1}\fi
\ifx \bbook \undefined \def \bbook#1{#1}\fi
\ifx \bcomment \undefined \def \bcomment#1{#1}\fi
\ifx \oauthor \undefined \def \oauthor#1{#1}\fi
\ifx \citeauthoryear \undefined \def \citeauthoryear#1{#1}\fi
\ifx \endbibitem  \undefined \def \endbibitem {}\fi
\ifx \bconflocation  \undefined \def \bconflocation#1{#1}\fi
\ifx \arxivurl  \undefined \def \arxivurl#1{\textsf{#1}}\fi
\csname PreBibitemsHook\endcsname

\bibitem[\protect\citeauthoryear{}{}]{nsdd}
\begin{botherref}
international network of Nuclear Structure and Decay Data evaluators (NSDD).
\url{https://www-nds.iaea.org/nsdd/}
\end{botherref}
\endbibitem

\bibitem[\protect\citeauthoryear{Hipp}{2020}]{sqlite}
\begin{botherref}
\oauthor{\bsnm{Hipp}, \binits{R.D.}}:
{SQLite}
(2020).
\url{https://www.sqlite.org/index.html}
\end{botherref}
\endbibitem

\bibitem[\protect\citeauthoryear{Kim et~al.}{2022}]{pubchem}
\begin{barticle}
\bauthor{\bsnm{Kim}, \binits{S.}},
\bauthor{\bsnm{Chen}, \binits{J.}},
\bauthor{\bsnm{Cheng}, \binits{T.}},
\bauthor{\bsnm{Gindulyte}, \binits{A.}},
\bauthor{\bsnm{He}, \binits{J.}},
\bauthor{\bsnm{He}, \binits{S.}},
\bauthor{\bsnm{Li}, \binits{Q.}},
\bauthor{\bsnm{Shoemaker}, \binits{B.A.}},
\bauthor{\bsnm{Thiessen}, \binits{P.A.}},
\bauthor{\bsnm{Yu}, \binits{B.}},
\bauthor{\bsnm{Zaslavsky}, \binits{L.}},
\bauthor{\bsnm{Zhang}, \binits{J.}},
\bauthor{\bsnm{Bolton}, \binits{E.E.}}:
\batitle{{PubChem 2023 update}}.
\bjtitle{Nucleic Acids Research}
\bvolume{51}(\bissue{D1}),
\bfpage{1373}--\blpage{1380}
(\byear{2022})
\doiurl{10.1093/nar/gkac956}
\end{barticle}
\endbibitem

\bibitem[\protect\citeauthoryear{Pezoa et~al.}{2016}]{json}
\begin{bchapter}
\bauthor{\bsnm{Pezoa}, \binits{F.}},
\bauthor{\bsnm{Reutter}, \binits{J.L.}},
\bauthor{\bsnm{Suarez}, \binits{F.}},
\bauthor{\bsnm{Ugarte}, \binits{M.}},
\bauthor{\bsnm{Vrgo{\v{c}}}, \binits{D.}}:
\bctitle{Foundations of json schema}.
In: \bbtitle{Proceedings of the 25th International Conference on World Wide
  Web},
pp. \bfpage{263}--\blpage{273}
(\byear{2016}).
\burl{https://dl.acm.org/doi/10.1145/2872427.2883029}
\end{bchapter}
\endbibitem

\bibitem[\protect\citeauthoryear{Thompson et~al.}{2009}]{xray-booklet}
\begin{botherref}
\oauthor{\bsnm{Thompson}, \binits{A.C.}},
\oauthor{\bsnm{Kirz}, \binits{J.}},
\oauthor{\bsnm{Attwood}, \binits{D.T.}},
\oauthor{\bsnm{Gullikson}, \binits{E.M.}},
\oauthor{\bsnm{Howells}, \binits{M.R.}},
\oauthor{\bsnm{Kortright}, \binits{J.B.}},
\oauthor{\bsnm{Liu}, \binits{Y.}},
\oauthor{\bsnm{Robinson}, \binits{A.L.}},
\oauthor{\bsnm{Underwood}, \binits{J.}},
\oauthor{\bsnm{Kim}, \binits{K.-J.}},
\oauthor{\bsnm{Lindau}, \binits{I.}},
\oauthor{\bsnm{Pianetta}, \binits{P.}},
\oauthor{\bsnm{Winick}, \binits{H.}},
\oauthor{\bsnm{Williams}, \binits{G.P.}},
\oauthor{\bsnm{Scofield}, \binits{J.H.}}:
X-Ray Data Booklet
(2009).
\url{https://cxro.lbl.gov//PDF/X-Ray-Data-Booklet.pdf}
\end{botherref}
\endbibitem

\bibitem[\protect\citeauthoryear{}{}]{ensdf}
\begin{botherref}
Evaluated Nuclear Structure Data File (ENSDF).
\url{https://www.nndc.bnl.gov/ensdf/}
\end{botherref}
\endbibitem

\bibitem[\protect\citeauthoryear{}{}]{NNDC}
\begin{botherref}
National Nuclear Data Center.
\url{https://www.nndc.bnl.gov/}
\end{botherref}
\endbibitem

\bibitem[\protect\citeauthoryear{}{}]{xundl}
\begin{botherref}
Experimental Unevaluated Nuclear Data List (XUNDL).
\url{https://www.nndc.bnl.gov/xundl/}
\end{botherref}
\endbibitem

\bibitem[\protect\citeauthoryear{}{}]{iaea-livechart}
\begin{botherref}
International Atomic Enzergy Agency: Live Chart of Nuclides.
\url{https://www-nds.iaea.org/relnsd/vcharthtml/VChartHTML.html}
\end{botherref}
\endbibitem

\bibitem[\protect\citeauthoryear{Huang et~al.}{2021}]{AME_2021}
\begin{barticle}
\bauthor{\bsnm{Huang}, \binits{W.J.}},
\bauthor{\bsnm{Wang}, \binits{M.}},
\bauthor{\bsnm{Kondev}, \binits{F.G.}},
\bauthor{\bsnm{Audi}, \binits{G.}},
\bauthor{\bsnm{Naimi}, \binits{S.}}:
\batitle{The ame 2020 atomic mass evaluation (i). evaluation of input data, and
  adjustment procedures}.
\bjtitle{Chinese Physics C}
\bvolume{45}(\bissue{3}),
\bfpage{030002}
(\byear{2021})
\doiurl{10.1088/1674-1137/abddb0}
\end{barticle}
\endbibitem

\bibitem[\protect\citeauthoryear{Wang et~al.}{2021}]{AME_2_2021}
\begin{barticle}
\bauthor{\bsnm{Wang}, \binits{M.}},
\bauthor{\bsnm{Huang}, \binits{W.J.}},
\bauthor{\bsnm{Kondev}, \binits{F.G.}},
\bauthor{\bsnm{Audi}, \binits{G.}},
\bauthor{\bsnm{Naimi}, \binits{S.}}:
\batitle{The ame 2020 atomic mass evaluation (ii). tables, graphs and
  references}.
\bjtitle{Chinese Physics C}
\bvolume{45}(\bissue{3}),
\bfpage{030003}
(\byear{2021})
\doiurl{10.1088/1674-1137/abddaf}
\end{barticle}
\endbibitem

\bibitem[\protect\citeauthoryear{Kondev et~al.}{2021}]{NUBASE_2021}
\begin{barticle}
\bauthor{\bsnm{Kondev}, \binits{F.G.}},
\bauthor{\bsnm{Wang}, \binits{M.}},
\bauthor{\bsnm{Huang}, \binits{W.J.}},
\bauthor{\bsnm{Naimi}, \binits{S.}},
\bauthor{\bsnm{Audi}, \binits{G.}}:
\batitle{The nubase2020 evaluation of nuclear physics properties *}.
\bjtitle{Chinese Physics C}
\bvolume{45}(\bissue{3}),
\bfpage{030001}
(\byear{2021})
\doiurl{10.1088/1674-1137/abddae}
\end{barticle}
\endbibitem

\bibitem[\protect\citeauthoryear{Angeli and Marinova}{2013}]{ANGELI201369}
\begin{barticle}
\bauthor{\bsnm{Angeli}, \binits{I.}},
\bauthor{\bsnm{Marinova}, \binits{K.P.}}:
\batitle{Table of experimental nuclear ground state charge radii: An update}.
\bjtitle{Atomic Data and Nuclear Data Tables}
\bvolume{99}(\bissue{1}),
\bfpage{69}--\blpage{95}
(\byear{2013})
\doiurl{10.1016/j.adt.2011.12.006}
\end{barticle}
\endbibitem

\bibitem[\protect\citeauthoryear{Stone}{2020}]{Stone_2020_Magnetic}
\begin{botherref}
\oauthor{\bsnm{Stone}, \binits{N.J.}}:
Table of Recommended Nuclear Magnetic Dipole Moments - Part II, Short-lived
  States.
INDC International Nuclear Data Committee
(2020).
\url{https://www-nds.iaea.org/publications/indc/indc-nds-0816.pdf}
\end{botherref}
\endbibitem

\bibitem[\protect\citeauthoryear{Stone}{2023a}]{Stone_2023_Magnetic}
\begin{botherref}
\oauthor{\bsnm{Stone}, \binits{N.J.}}:
Table of Recommended Nuclear Magnetic Dipole Moments: Part I - Long-lived
  States.
INDC International Nuclear Data Committee
(2023).
\doiurl{10.61092/iaea.yjpc-cns6} .
\url{https://doi.org/10.61092/iaea.yjpc-cns6}
\end{botherref}
\endbibitem

\bibitem[\protect\citeauthoryear{Stone}{2023b}]{Stone_2023_Electric}
\begin{botherref}
\oauthor{\bsnm{Stone}, \binits{N.J.}}:
Table of Nuclear Electric Quadrupole Moments.
INDC International Nuclear Data Committee
(2023).
\url{https://www-nds.iaea.org/publications/indc/indc-nds-0833.pdf}
\end{botherref}
\endbibitem

\bibitem[\protect\citeauthoryear{}{}]{iaea-api}
\begin{botherref}
IAEA LiveChart API.
\url{https://www-nds.iaea.org/relnsd/vcharthtml/api_v0_guide.html}
\end{botherref}
\endbibitem

\bibitem[\protect\citeauthoryear{}{}]{nudat}
\begin{botherref}
National Nuclear Data Center, information extracted from the NuDat database.
\url{https://www.nndc.bnl.gov/nudat/}
\end{botherref}
\endbibitem

\bibitem[\protect\citeauthoryear{Project}{2008}]{crontab}
\begin{botherref}
\oauthor{\bsnm{Project}, \binits{T.L.D.}}:
Crontab(1) - Linux Man Page.
Linux Online,
(2008).
Linux Online.
\url{http://manpages.ubuntu.com/manpages/trusty/man1/crontab.1.html}
\end{botherref}
\endbibitem

\bibitem[\protect\citeauthoryear{Tuli}{2001}]{ensdf-format}
\begin{botherref}
\oauthor{\bsnm{Tuli}, \binits{J.K.}}:
The Evaluated Nuclear Structure Data File. A Manual for Preparation of Data
  Sets
(2001).
\url{https://www-nds.iaea.org/public/documents/ensdf/}
\end{botherref}
\endbibitem

\bibitem[\protect\citeauthoryear{}{}]{cmake}
\begin{botherref}
CMake.
\url{https://cmake.org}
\end{botherref}
\endbibitem

\bibitem[\protect\citeauthoryear{Brun and Rademakers}{1997}]{root}
\begin{barticle}
\bauthor{\bsnm{Brun}, \binits{R.}},
\bauthor{\bsnm{Rademakers}, \binits{F.}}:
\batitle{Root — an object oriented data analysis framework}.
\bjtitle{Nucl. Inst. and Meth. A}
\bvolume{389}(\bissue{1}),
\bfpage{81}--\blpage{86}
(\byear{1997})
\doiurl{10.1016/S0168-9002(97)00048-X}
\end{barticle}
\endbibitem

\bibitem[\protect\citeauthoryear{Lohmann}{1970}]{lohmann}
\begin{botherref}
\oauthor{\bsnm{Lohmann}, \binits{N.}}:
JSON for Modern C++.
Zenodo
(1970).
\doiurl{10.5281/zenodo.5814096}
\end{botherref}
\endbibitem

\bibitem[\protect\citeauthoryear{Dudouet and Gruyer}{}]{tkn-git}
\begin{botherref}
\oauthor{\bsnm{Dudouet}, \binits{J.}},
\oauthor{\bsnm{Gruyer}, \binits{D.}}:
TkN library git repository.
\url{https://gitlab.in2p3.fr/tkn/tkn-lib}
\end{botherref}
\endbibitem

\bibitem[\protect\citeauthoryear{}{}]{cecill-b}
\begin{botherref}
CeCILL-B free software licensed agreement.
\url{https://cecill.info/licences/Licence_CeCILL-B_V1-en.html}
\end{botherref}
\endbibitem

\bibitem[\protect\citeauthoryear{Merkel}{2014}]{docker}
\begin{barticle}
\bauthor{\bsnm{Merkel}, \binits{D.}}:
\batitle{Docker: lightweight linux containers for consistent development and
  deployment}.
\bjtitle{Linux journal}
\bvolume{2014}(\bissue{239}),
\bfpage{2}
(\byear{2014})
\doiurl{10.5555/2600239.2600241}
\end{barticle}
\endbibitem

\bibitem[\protect\citeauthoryear{}{}]{googletest}
\begin{botherref}
GoogleTest.
\url{https://github.com/google/googletest}
\end{botherref}
\endbibitem

\bibitem[\protect\citeauthoryear{}{}]{cppcheck}
\begin{botherref}
Cppcheck.
\url{https://cppcheck.sourceforge.io}
\end{botherref}
\endbibitem

\bibitem[\protect\citeauthoryear{}{}]{sonarqube}
\begin{botherref}
SonarQube.
\url{https://www.sonarsource.com/products/sonarqube}
\end{botherref}
\endbibitem

\bibitem[\protect\citeauthoryear{Dudouet and Gruyer}{}]{tkn-doc}
\begin{botherref}
\oauthor{\bsnm{Dudouet}, \binits{J.}},
\oauthor{\bsnm{Gruyer}, \binits{D.}}:
TkN library documentation.
\url{https://tkn.in2p3.fr}
\end{botherref}
\endbibitem

\bibitem[\protect\citeauthoryear{van Heesch}{}]{doxygen}
\begin{botherref}
\oauthor{\bsnm{Heesch}, \binits{D.}}:
Doxygen: Documentation Generator.
\url{http://www.doxygen.nl/}
\end{botherref}
\endbibitem

\bibitem[\protect\citeauthoryear{Dudouet}{}]{cubix-userguide}
\begin{botherref}
\oauthor{\bsnm{Dudouet}, \binits{J.}}:
The Cubix spectra viewer.
\url{https://cubix.in2p3.fr/}
\end{botherref}
\endbibitem

\bibitem[\protect\citeauthoryear{Otuka et~al.}{2014}]{exfor}
\begin{barticle}
\bauthor{\bsnm{Otuka}, \binits{N.}},
\bauthor{\bsnm{Dupont}, \binits{E.}},
\bauthor{\bsnm{Semkova}, \binits{V.}},
\bauthor{\bsnm{Pritychenko}, \binits{B.}},
\bauthor{\bsnm{Blokhin}, \binits{A.I.}},
\bauthor{\bsnm{Aikawa}, \binits{M.}},
\bauthor{\bsnm{Babykina}, \binits{S.}},
\bauthor{\bsnm{Bossant}, \binits{M.}},
\bauthor{\bsnm{Chen}, \binits{G.}},
\bauthor{\bsnm{Dunaeva}, \binits{S.}},
\bauthor{\bsnm{Forrest}, \binits{R.A.}},
\bauthor{\bsnm{Fukahori}, \binits{T.}},
\bauthor{\bsnm{Furutachi}, \binits{N.}},
\bauthor{\bsnm{Ganesan}, \binits{S.}},
\bauthor{\bsnm{Ge}, \binits{Z.}},
\bauthor{\bsnm{Gritzay}, \binits{O.O.}},
\bauthor{\bsnm{Herman}, \binits{M.}},
\bauthor{\bsnm{Hlavač}, \binits{S.}},
\bauthor{\bsnm{Katō}, \binits{K.}},
\bauthor{\bsnm{Lalremruata}, \binits{B.}},
\bauthor{\bsnm{Lee}, \binits{Y.O.}},
\bauthor{\bsnm{Makinaga}, \binits{A.}},
\bauthor{\bsnm{Matsumoto}, \binits{K.}},
\bauthor{\bsnm{Mikhaylyukova}, \binits{M.}},
\bauthor{\bsnm{Pikulina}, \binits{G.}},
\bauthor{\bsnm{Pronyaev}, \binits{V.G.}},
\bauthor{\bsnm{Saxena}, \binits{A.}},
\bauthor{\bsnm{Schwerer}, \binits{O.}},
\bauthor{\bsnm{Simakov}, \binits{S.P.}},
\bauthor{\bsnm{Soppera}, \binits{N.}},
\bauthor{\bsnm{Suzuki}, \binits{R.}},
\bauthor{\bsnm{Takács}, \binits{S.}},
\bauthor{\bsnm{Tao}, \binits{X.}},
\bauthor{\bsnm{Taova}, \binits{S.}},
\bauthor{\bsnm{Tárkányi}, \binits{F.}},
\bauthor{\bsnm{Varlamov}, \binits{V.V.}},
\bauthor{\bsnm{Wang}, \binits{J.}},
\bauthor{\bsnm{Yang}, \binits{S.C.}},
\bauthor{\bsnm{Zerkin}, \binits{V.}},
\bauthor{\bsnm{Zhuang}, \binits{Y.}}:
\batitle{Towards a more complete and accurate experimental nuclear reaction
  data library (exfor): International collaboration between nuclear reaction
  data centres (nrdc)}.
\bjtitle{Nuclear Data Sheets}
\bvolume{120},
\bfpage{272}--\blpage{276}
(\byear{2014})
\doiurl{10.1016/j.nds.2014.07.065}
\end{barticle}
\endbibitem

\bibitem[\protect\citeauthoryear{}{}]{exfor-web}
\begin{botherref}
Experimental Nuclear Reaction Data (EXFOR).
\url{https://www-nds.iaea.org/exfor/}
\end{botherref}
\endbibitem

\end{thebibliography}

\begin{appendices}

\onecolumn

\section{Database content} \label{app-tables}

\begin{table}[ht]
\begin{tabular}{r c c c}
\hline
Property  & Uncertainty & Unit  & Source \\
\hline
proton number &  &  & \cite{pubchem} \\
name (text: e.g. Helium, Gold...) &  &  & \cite{pubchem} \\
symbol (text: e.g. He, Au...) &  &  & \cite{pubchem} \\
atomic mass &  & u  & \cite{pubchem} \\
atomic radius van der Waals &  & $pm$  & \cite{pubchem} \\
boiling point &  & K  & \cite{pubchem} \\
density &  & $g/cm^3$ & \cite{pubchem} \\
electronic configuration (text) &  &  & \cite{pubchem} \\
group block (text: e.g. metal, alkaline...) &  &  & \cite{pubchem} \\
ionization energy &  & eV  & \cite{pubchem} \\
melting point &  & K  & \cite{pubchem} \\
standard state (text: e.g. solid, liquid...) &  &  & \cite{pubchem} \\
year discovered (text) &  &  & \cite{pubchem} \\
$K\alpha_{1}$ X-ray energy &  & keV & \cite{xray-booklet} \\
$K\alpha_{2}$ X-ray energy &  & keV & \cite{xray-booklet} \\
$K\beta_{1}$ X-ray energy &  & keV & \cite{xray-booklet} \\
$L\alpha_{1}$ X-ray energy &  & keV & \cite{xray-booklet} \\
$L\alpha_{2}$ X-ray energy &  & keV & \cite{xray-booklet} \\
$L\beta_{1}$ X-ray energy &  & keV & \cite{xray-booklet} \\
$L\beta_{2}$ X-ray energy &  & keV & \cite{xray-booklet} \\
$L\gamma_{1}$ X-ray energy &  & keV & \cite{xray-booklet} \\
$M\alpha_{1}$ X-ray energy &  & keV & \cite{xray-booklet} \\
\hline
\end{tabular}
\caption{\label{table-element} Content of the element table.}
\end{table}

\begin{table}[ht]
\begin{tabular}{r c c c}
\hline
Property  & Uncertainty & Default unit  & Source \\
\hline
nucleon number &  &  & \cite{iaea-api} \\
lifetime & \checkmark  & adaptive  & \cite{ensdf,xundl} \\
abundance & \checkmark  & \%  & \cite{nudat} \\
decay modes (text) &  &  & \cite{nudat} \\
spin parity (text) &  &  & \cite{ensdf,xundl} \\
$\beta_{2}$: quadrupole deformation parameter & \checkmark  &   & \cite{nudat} \\
mass excess & \checkmark  & keV  & \cite{iaea-api} \\
binding energy per nucleon & \checkmark  & keV  & \cite{iaea-api}$^{1}$ \\
pairing gap & \checkmark  & keV  & \cite{iaea-api}$^{1}$ \\
$S_{n}$: neutron separation energy & \checkmark  & keV  & \cite{iaea-api}$^{1}$ \\
$S_{p}$: proton separation energy & \checkmark  & keV  & \cite{iaea-api}$^{1}$ \\
$S_{2n}$: two neutron separation energy& \checkmark  & keV  & \cite{iaea-api}$^{1}$ \\
$S_{2p}$: two proton separation energy & \checkmark  & keV  & \cite{iaea-api}$^{1}$ \\
$Q_{\alpha}$: alpha Q-value & \checkmark  & keV  & \cite{iaea-api}$^{1}$ \\
$Q_{\beta^{-}}$: $\beta^{-}$ Q-value & \checkmark  & keV  & \cite{iaea-api}$^{1}$ \\
$Q_{\beta^{-}-n}$: beta-delayed neutron emission Q-value & \checkmark  & keV  & \cite{iaea-api}$^{1}$ \\
$Q_{\beta^{-}-2n}$: beta-delayed double neutron emission Q-value & \checkmark  & keV  & \cite{iaea-api}$^{1}$ \\
$\Delta Q_{\alpha}$: $0.5\times( Q_{\alpha}(Z+2,N+2) - Q_{\alpha}(Z,N))$ & \checkmark  & keV  & \cite{iaea-api}$^{1}$ \\
$Q_{2\beta^{-}}$:double $\beta^{-}$ Q-value & \checkmark  & keV  & \cite{iaea-api}$^{1}$ \\
$Q_{2EC}$: double electron capture Q-value & \checkmark  & keV  & \cite{iaea-api}$^{1}$ \\
$Q_{EC}$: electron capture Q-value & \checkmark  & keV  & \cite{iaea-api}$^{1}$ \\
$Q_{EC-p}$:electron capture followed by proton emission Q-value & \checkmark  & keV  & \cite{iaea-api}$^{1}$ \\
$Q_{\beta^{+}}$: positron emission Q-value & \checkmark  & keV  & \cite{iaea-api}$^{1}$ \\
thermal neutron-induced fission yields for $^{235}$U & \checkmark  & normalized to 2  & \cite{nudat} \\
thermal neutron-induced fission yields for $^{239}$Pu & \checkmark  & normalized to 2  & \cite{nudat} \\
spontaneous fission yields for $^{252}$Cf & \checkmark  & normalized to 2  & \cite{nudat} \\
\hline
\end{tabular}
\caption{\label{table-isotope} Content of the isotope table. Sources labeled as \cite{iaea-api}$^{1}$ are not directly taken from \cite{iaea-api} but re-calculated in TkN from the mass-excess value.}
\end{table}

\begin{table}[ht]
\begin{tabular}{r c c c}
\hline
Property  & Uncertainty & Default unit  & Source \\
\hline
name (text) &  &  & \cite{ensdf,xundl} \\
comment (text) &  &  & \cite{ensdf,xundl} \\
\hline
\end{tabular}
\caption{\label{table-dataset} Content of the dataset table used to access to the global datasets comments specified by NSDD evaluators.}
\end{table}

\begin{table}[ht]
\begin{tabular}{r c c c}
\hline
Property  & Uncertainty & Default unit  & Source \\
\hline
energy & \checkmark  & keV  & \cite{ensdf,xundl} \\
half-life & \checkmark  & adaptive  & \cite{ensdf,xundl} \\
spin parity (text) &  &  & \cite{ensdf,xundl} \\
comment (text) &  &  & \cite{ensdf,xundl} \\
\hline
\end{tabular}
\caption{\label{table-level} Content of the level table.}
\end{table}

\begin{table}[ht]
\begin{tabular}{r c c c}
\hline
Name  & Uncertainty & Default unit  & Source \\
\hline
energy & \checkmark  & keV  & \cite{ensdf,xundl} \\
relative intensity & \checkmark  & \%  & \cite{ensdf,xundl} \\
multipolarity of transition (text)  &  &  & \cite{ensdf,xundl} \\
total conversion coefficient & \checkmark  &   & \cite{ensdf,xundl} \\
mixing ratio $\delta$ & \checkmark  &  & \cite{ensdf,xundl} \\
Reduced electric transition probability (downward) & \checkmark  & $e^2\times(fm)^L$  & \cite{ensdf,xundl} \\
Reduced electric transition probability & \checkmark  & Weisskopf  & \cite{ensdf,xundl} \\
Reduced magnetic transition probability (downward) & \checkmark  & $\mu^2 \times (fm)^{L-1}$  & \cite{ensdf,xundl} \\
Reduced magnetic transition probability & \checkmark  & Weisskopf  & \cite{ensdf,xundl} \\
comment (text) &  &  & \cite{ensdf,xundl} \\
\hline
\end{tabular}
\caption{\label{table-decay} Content of the decay table.}
\end{table}

\end{appendices}

\end{document}